\documentclass[twocolumn, superscriptaddress, prl, showpacs, footinbib]{revtex4-1}

\usepackage{graphicx}
\usepackage{multirow}

\begin{document}

\title{Diffusion of Small Molecules in Metal Organic Framework Materials}

\author{Pieremanuele Canepa}
\affiliation{Department of Physics, Wake Forest University,
Winston-Salem, NC 27109, USA.}

\author{Nour Nijem}
\affiliation{Department of Materials Science and Engineering, University
of Texas at Dallas, TX 75080, USA.}

\author{Yves J. Chabal}
\affiliation{Department of Materials Science and Engineering, University
of Texas at Dallas, TX 75080, USA.}

\author{T. Thonhauser}
\email[E-mail: ]{thonhauser@wfu.edu}
\affiliation{Department of Physics, Wake Forest University,
Winston-Salem, NC 27109, USA.}

\date{\today}

\begin{abstract}
\emph{Ab initio} simulations are combined with \emph{in situ} infrared
spectroscopy to unveil the molecular transport of H$_2$, CO$_2$, and
H$_2$O in the metal organic framework MOF-74-Mg.  Our study
uncovers---at the atomistic level---the major factors governing the
transport mechanism of these small molecules. In particular, we identify
four key diffusion mechanisms and calculate the corresponding diffusion
barriers, which are nicely confirmed by time-resolved infrared
experiments. We also answer a long-standing question about the existence
of secondary adsorption sites for the guest molecules, and we show how
those sites affect the macroscopic diffusion properties.  Our findings
are important to gain a fundamental understanding of the diffusion
processes in these nano-porous materials, with direct implications for
the usability of MOFs in gas sequestration and storage applications.
\end{abstract}

\maketitle

Distressing scenarios of fossil-fuel shortages and greenhouse effects
have intensified scientific efforts in seeking innovative materials for
energy storage and capture of greenhouse gases.  Due to their vast range
of tunable properties, metal organic framework (MOF) materials are
outstanding candidates to address both issues. They can have large
hydrogen-storage capacities, surpassing the US DOE storage
targets~\cite{Furukawa07, Murray09, new_targets}, and have even been
declared DOE's top priority at its \emph{2012 Hydrogen and Fuel Cells
Program Annual Review Meeting}~\cite{ARM}. On the other hand, due to
their large CO$_2$ uptakes~\cite{Bao11, Britt09}, MOFs are also
interesting for carbon-capture applications.  These qualities, combined
with their low production cost, have made MOFs the target of many
studies, focusing mostly on their adsorption properties~\cite{Yang10,
new_targets, Murray09, Li09, Li11}.  However, their performance for
practical storage and capture applications also critically depends on
the diffusion and interaction of gases (e.g. with water) in the MOF
environment, which are currently poorly understood~\cite{Britt08,
Britt09, Glover11}.

To address this issue and elucidate the diffusion process of small
molecules in the nano-pores of MOFs, we use a combination of \emph{ab
initio} simulations and \emph{in situ} infrared (IR) spectroscopy.
Specifically, we study the diffusion of H$_2$, CO$_2$, and H$_2$O in
MOF-74-Mg and investigate the role of water on diffusion of small
molecules.  We focus on this particular MOF, as it has attracted a lot
of attention due to its enhanced reactivity with small molecules, caused
by the exposure of open metal sites in its nano-pores.  Britt \emph{et
al.}~\cite{Britt09} have shown that MOF-74-Mg is extremely efficient in
capturing CO$_2$, compared to iso-structural MOFs with other metal sites
such as Zn, Mn, Fe, Co, Ni, and Cu.

To model the molecular diffusion in the MOF structure, we use
climbing-image nudged elastic band~\cite{Henkelman00} (NEB) simulations
coupled with density functional theory (DFT), as implemented in
\textsc{QuantumEspresso}~\cite{Giannozzi09}. To correctly capture the
weak van der Waals forces---which are critical in this study---we use
the truly non-local functional vdW-DF~\cite{Dion07, Thonhauser07,
Langreth09}, which has already been successfully applied to a number of
related studies~\cite{Yao12,Kong09,Kong09B,Kong11,Perez10}. We use
ultrasoft pseudopotentials with wave-function and density cutoffs of
$480$ eV and $4800$ eV, respectively. Tests show that $\Gamma$-point
calculations are sufficient and yield total energies converged to within
5~meV with respect to denser \emph{k}-point meshes; however, energy
\emph{differences}---important for our diffusion barriers---are
converged to within less than 1~meV.  The self-consistency tolerance was
set to $1.4\times10^{-10}$ eV and during optimizations the total forces
were relaxed to less than $2.6\times10^{-4}$~eV/\AA. We started from the
experimental MOF-74-Mg structure, optimizing the internal parameters and
keeping the lattice constants fixed to the hexagonal structure with
space group $R\overline{3}$ and $a=25.8810$~\AA\ and $c =
6.8789$~\AA{}~\cite{Wu09}.

\begin{figure}
\begin{center}
\includegraphics[width=0.67\columnwidth]{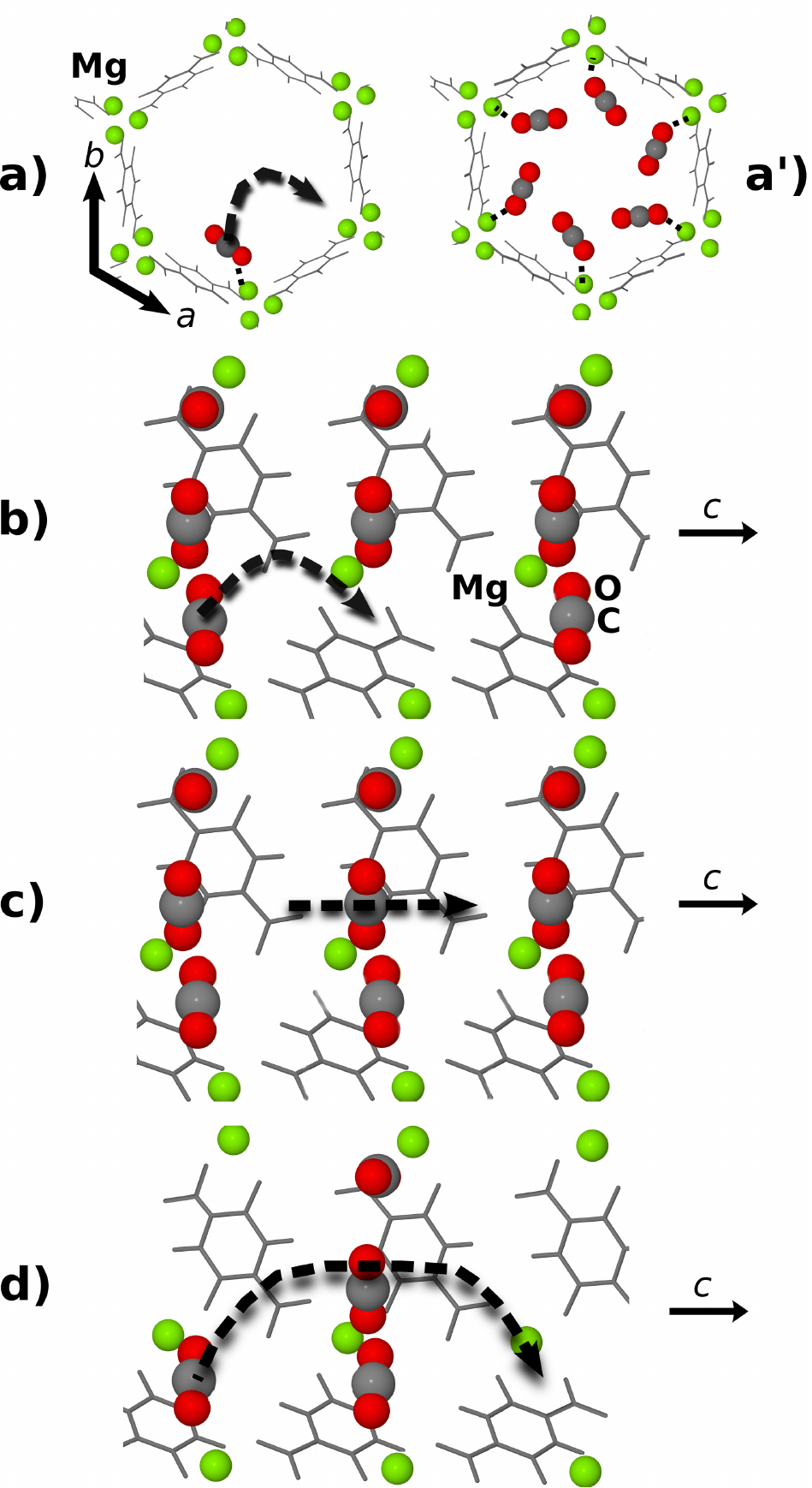}
\end{center}
\caption{\label{fig:1_schema} Graphical representation of the diffusion
mechanisms considered in this study, shown for the case of CO$_2$.  a)
and a') are views directly along the $c$-axis of the hexagonal MOF-74-Mg
cell, where one (low loading) and six CO$_2$ (high loading) are
adsorbed. b), c), and d) are views perpendicular to the \emph{c}-axis.
Dashed lines indicate the diffusion paths.}
\end{figure}

We begin by revisiting the binding characteristics of H$_2$, CO$_2$, and
H$_2$O in the MOF framework, which serves here as background for
explaining the molecular transport---recent
theoretical~\cite{Valenzano10, Valenzano11, Nijem12} and
experimental~\cite{Valenzano10, Nijem12, Nijem10, Zhou08, Queen11} work
has fully explored and clarified both origin and nature of the
``static'' adsorption interactions of H$_2$ and CO$_2$ in MOF-74. Our
calculated adsorption energies $\Delta E$ are given in
Table~\ref{table:be} for two different loadings: (i) low-loading, one
molecule per cell, and (ii) high-loading, six molecules per cell
completely saturating all available metal sites (see
Fig.~\ref{fig:1_schema}a and Fig.~\ref{fig:1_schema}a' for a graphical
representation of these loadings).  From the computed vibrational
frequencies (see below) we obtain the thermal and zero-point energy
(ZPE) corrections to the adsorption energy, allowing for a more accurate
comparison to measured adsorption heats. But, we find that neither of
these corrections has a large effect in our case. Overall, we find very
good agreement with the experimental adsorption energies of --0.11 $\pm$
0.003 eV for H$_2$~\cite{Zhou08} and --0.49 $\pm$ 0.010 eV for
CO$_2$~\cite{Valenzano10}, attesting to the importance of correctly
including van der Waals interactions in these simulations. The 0.03~eV
decrease of $\Delta E$ for H$_2$O in the high-loading situation is
linked to the intermolecular repulsions between hydrogen atoms of H$_2$O
adsorbed on adjacent Mg sites. This is also evident from the
intermolecular distance $d_{\rm HHO\cdots Mg-MOF}$, which increases from
2.197~\AA\, for the low-loading case to 2.232~\AA\, for the high-loading
case. As a result of the large dipole moment of water (1.9~Debye), it
binds very strongly~\cite{Murray09}.  Thus, water is thermodynamically
much more likely to occupy metal sites, significantly reducing the MOF's
adsorption capabilities towards H$_2$ and CO$_2$, with tremendous
importance for storage and capture applications.

\begin{table}
\caption{\label{table:be}Adsorption energy $\Delta E$ of molecules in
MOF-74-Mg in eV, and adsorption energies corrected for the zero-point
energy ($\Delta E_{\rm ZPE}$) and thermal contribution ($\Delta H_{\rm
298}$ at 298 K).}
\begin{tabular*}{\columnwidth}{@{\extracolsep{\fill}}lcccr@{}}\hline\hline
Molecule  & Loading & $\Delta E$ & $\Delta E_{\rm ZPE}$  & $\Delta H_{\rm 298}$\\\hline
\multirow{2}{*}{H$_2$}  & 1 & --0.15 & --0.15  & --0.15\\
                        & 6 & --0.16 & --0.16  & --0.16\\
\multirow{2}{*}{CO$_2$} & 1 & --0.50 & --0.49  & --0.50\\
                        & 6 & --0.50 & --0.49  & --0.50\\
\multirow{2}{*}{H$_2$O} & 1 & --0.79 & --0.76  & --0.76\\
                        & 6 & --0.76 & --0.73  & --0.73\\\hline\hline
\end{tabular*}
\end{table}

While the uncoordinated Mg sites are the most attractive sites,
secondary binding sites also exist, which only become occupied at lower
temperatures (or high pressures). Recent neutron diffraction experiments
isolated such a secondary binding site for CO$_2$ in MOF-74-Mg
\cite{Queen11}, where the CO$_2$ binds to the carboxylate group of the
linker. The experimental geometry of this site is well reproduced by our
vdW-DF calculations \cite{Nijem12}. For very high loading, i.e.\ 12
CO$_2$ molecules per unit cell (occupying the primary and secondary
sites) the inter-molecular repulsions decrease the average $\Delta E$ to
--0.48~eV.

Vibrational frequencies are a prerequisite to estimate the
pre-exponential factor in the Arrhenius transport equation. We report
here the calculated \emph{change} in IR frequency between the gas-phase
molecules and the molecules adsorbed in the MOF. In the mono-adsorbed
cases we find $\Delta \nu_{\rm H_2}$ = --30 cm$^{-1}$, $\Delta \nu_{\rm
CO_2}$ = --13 cm$^{-1}$, and $\Delta \nu_{\rm H_2O}$ = --103 cm$^{-1}$,
in excellent agreement with our experimental observations of --36
cm$^{-1}$, --8 cm$^{-1}$ and --99~cm$^{-1}$.  Experimentally, there is a
small difference in frequency change between low and high loading,
resulting in a red-shift of --3 cm$^{-1}$ and --15~cm$^{-1}$ for the
asymmetric stretch mode of CO$_2$ and H$_2$O (see supplementary
material). Computationally, we find $\Delta \nu_{\rm CO_2}$ = --1
cm$^{-1}$ and $\Delta \nu_{\rm H_2O}$ = --18 cm$^{-1}$, in excellent
agreement with experiment. 

We now move our discussion to the key results of this letter: The
molecular transport in MOF-74-Mg. To study this aspect, we consider the
four diffusion mechanisms depicted in Fig.~\ref{fig:1_schema}. In
mechanism a), a molecule hops circularly from one Mg$^{2+}$ site to its
adjacent one.  Note that this mechanism is not responsible for molecular
transport \emph{into} the MOF, but nevertheless it is an important
process for redistributing the load. In mechanism b), one molecule hops
longitudinally (along the \emph{c}-axis) from one Mg$^{2+}$ to the
equivalent one in the next unit cell.  In mechanism c), one molecule
moves longitudinally through MOF-74 fully loaded with the same type of
molecule. And finally, in mechanism d), one molecule pre-adsorbed on a
Mg$^{2+}$ site moves through a barrier made by six molecules and then
binds again at the equivalent site two unit cells further down. We
consider those the fundamental diffusion mechanisms that control the
macroscopic molecular transport in MOF-74-Mg. Other paths by good
approximation are superpositions of the ones discussed here.  Mechanism
c) simulates real diffusion and is responsible for penetration deep into
the MOF, once the surface is fully saturated. Mechanism d) simulates the
kinetic barrier that a pre-adsorbed molecule must overcome when the
adjacent available metals sites have already been saturated, i.e.\
obstructing the flux of an incoming molecule. Note that co-diffusion
cases are not considered here.  For the study of mechanism a) we used
the hexagonal primitive cell with 54 atoms. For mechanism b), c), and
d)---requiring a longitudinal displacement of the molecule along the
$c$-axis---we employed a supercell containing 108 atoms, expanded along
the $c$-axis, making these calculations very challenging.

\begin{figure}
\begin{center}
\includegraphics[width=\columnwidth]{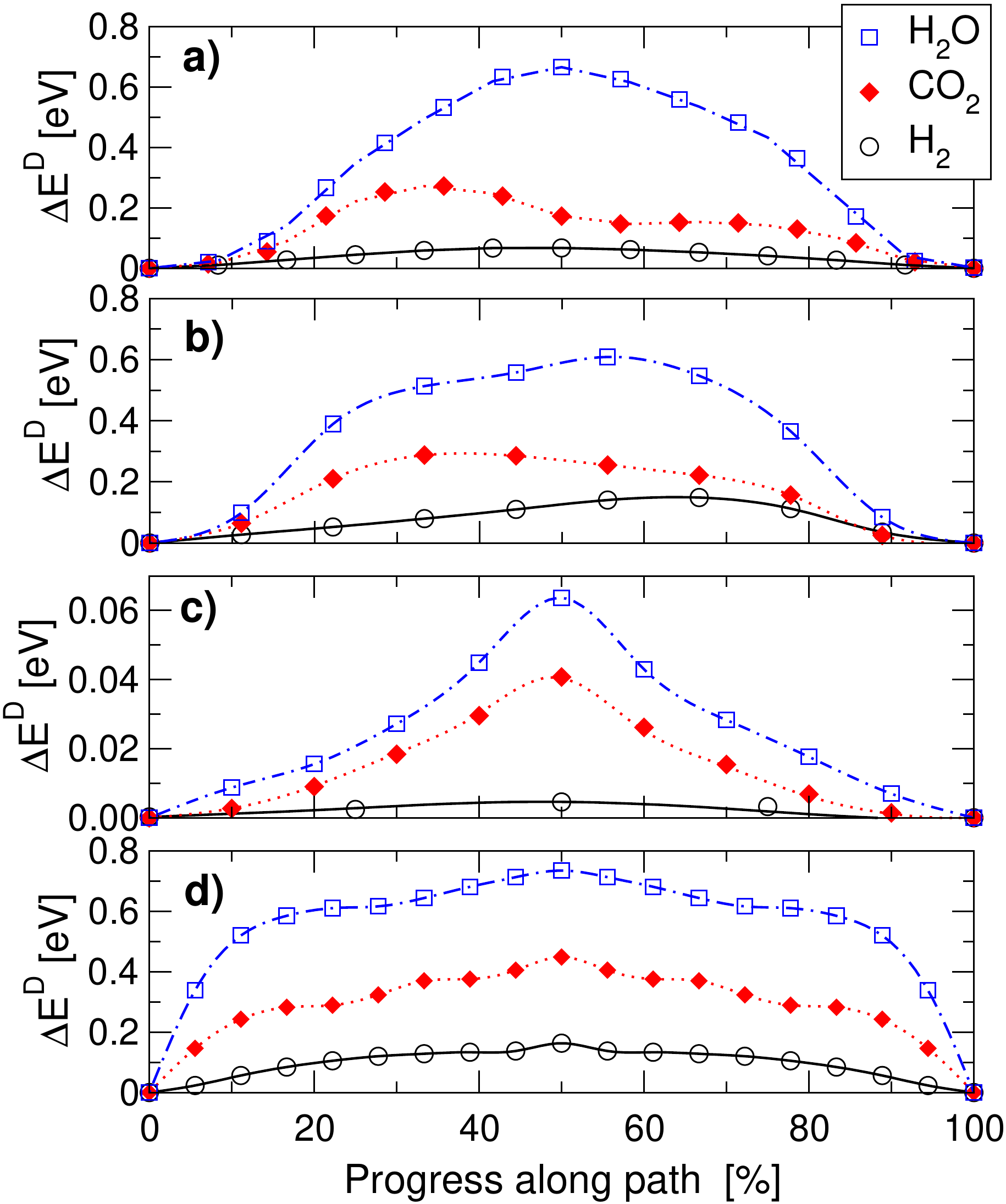}
\end{center}
\caption{\label{fig:neb} Diffusion profiles (in eV) for the diffusion
processes of H$_2$, CO$_2$, and H$_2$O in MOF-74-Mg according to the
mechanisms in Fig~\ref{fig:1_schema}.}
\end{figure}

\begin{table}
\caption{\label{table:bar} Diffusion barriers $\Delta E^D$ (in eV) of
small molecules in MOF-74-Mg. ZPE corrected barriers $\Delta E^D_{\rm
ZPE}$ are given in brackets.}
\begin{tabular*}{\columnwidth}{@{\extracolsep{\fill}}lcccc@{}}\hline\hline
       & \multicolumn{4}{c}{diffusion mechanism}\\
       &  a) & b) & c) & d)\\\hline
H$_2$  &  0.07 (0.07)  &  0.15 (0.15) & 0.005 (0.004) & 0.16 (0.16)\\
CO$_2$ &  0.27 (0.26)  &  0.29 (0.28) & 0.04  (0.03)  & 0.45 (0.44)\\
H$_2$O &  0.67 (0.64)  &  0.61 (0.58) & 0.06  (0.04)  & 0.73 (0.70)\\\hline \hline
\end{tabular*}
\end{table}

Figure~\ref{fig:neb}, consistently labeled with Fig.~\ref{fig:1_schema},
shows the calculated diffusion barriers for H$_2$, CO$_2$, and H$_2$O.
The barrier heights of all curves, i.e.\ their maxima, will be referred
to as $\Delta E^D$ and are collected in Table~\ref{table:bar}. Not
surprisingly, looking at the circular hopping mechanism a), we see that
the molecules experience diffusion barriers similar in magnitudes to the
adsorption energies at the metal site. For both H$_2$ and H$_2$O the
diffusion barriers look alike and are found very symmetric. This is not
the case for CO$_2$, which shows a second minimum located at 58\% (see
Fig.~\ref{fig:neb}a). The local geometry of this minimum is that of the
secondary adsorption site detected by neutron diffraction in
Ref.~\onlinecite{Queen11}. Figure~\ref{fig:neb}a) shows that the
secondary adsorption site becomes only activated when all the six
Mg$^{2+}$ sites are fully occupied, as the local minimum (5~meV) is too
shallow to trap molecules even at very low temperatures. In other words,
the secondary binding site only gets occupied under very high
loading/pressure.  Barriers calculated for mechanism b) are similar in
magnitude to those of mechanism a), however, they are slightly more
asymmetric. We considered mechanism b) also for situations of
low-loading (not shown in Fig.~\ref{fig:neb}), i.e.\ only one molecule
occupies the unit cell and hops along the $c$-axis. In this case, all
barriers remain similar with the exception of that for water, which
increases by approximately 10\%, suggesting that the crowded environment
helps the molecule extraction via an expulsion mechanism. Not
coincidentally, the maximum of the H$_2$O barrier of
Fig.~\ref{fig:neb}b) corresponds to the formation-breaking of a lateral
hydrogen-bond with an oxygen nearby the Mg sites (OOH$\cdots$O--MOF),
which decreases its strength passing from low ($d_{\rm OOH \cdots
O-MOF}$ = 2.077~\AA) to high loadings ($d_{\rm OOH\cdots O-MOF}$ = 2.138~\AA).
It follows that H$_2$O experiences the largest diffusion barriers and has
the highest adsorption affinity to metal sites (see
Table~\ref{table:be}).

Although not immediately obvious from Fig.~\ref{fig:neb}b), for H$_2$
there is a small secondary minimum near the 50\% mark, which is off the
optimized pathway and thus does not show up in the NEB calculation.
With the help of additional optimizations near that point, we confirm
the presence of a secondary binding site with an adsorption energy of
$\Delta E = -0.11$~eV, which is 0.04~eV above the primary binding site
(see Table~\ref{table:be}).  The binding pocket is relatively shallow
with a depth of approximately 2~meV and the H$_2$ is clearly ``docked''
on the ``oxygen triangle'' formed by the linkers coordinating the Mg
sites.  Evidence of this binding site was also found in MOF-74-Zn both
theoretically~\cite{Kong09} and with neutron diffraction
experiments~\cite{Liu08}. Our calculations thus confirm and quantify the
schematic kinetic model (based on experimental evidences) proposed in
Fig.~12 of Ref.~\cite{Nijem10}, where the gas-phase H$_2$ molecules
first condense on the metastable second binding site, followed by a
gradual migration to the more stable Mg site. Also note that the
asymmetric profile for this mechanism suggests that the release of a
H$_2$ molecule from the Mg site is a direct process without intermediate
steps.  From this analysis we learn that at high pressure and
temperature the existence of secondary binding sites assists in the
quick redistribution of the molecular load, as molecules entrapped in
these sites need less energy to escape from their minima.  These sites
thus play a crucial role in the load redistribution, affecting the final
diffusion properties of incoming molecules. For example, when water is
adsorbed at the primary and secondary sites, we can imagine scenarios of
high hydrogen bonding reticulation, severely limiting the molecular
transport of H$_2$ and CO$_2$.

Next, we focus on mechanism c) and d), which aim to simulate the
molecular flow in the MOF. In mechanism c), we calculate the barrier for
diffusion of a single molecule through a MOF fully loaded with the same
type of molecule.  Again, water suffers the largest impediment to flow
(0.06~eV), while H$_2$ for this mechanism is almost barrier-free
(5~meV).  Just recently, Bao \emph{et al.}~\cite{Bao11} measured $\Delta
E^D$ for CO$_2$ and found 0.03~eV, based on experimental CO$_2$
adsorption isotherms. This is in perfect agreement with our computed
values of 0.04~eV and 0.03~eV (ZPE corrected). Mechanism d) considers a
situation where an adsorbed molecule tries to diffuse further into the
MOF, hampered by the presence of a molecular barrier due to other
pre-adsorbed molecules. In this case, the strong electrostatic
interaction established by water with the Mg sites imposes a larger
diffusion barrier for H$_2$O when compared to CO$_2$ and H$_2$.

We now correlate our diffusion mechanisms at the atomistic level
discussed above with our \emph{in situ} time-resolved IR diffusion
experiments, which provide information about the macroscopic diffusion
of molecules within the MOF scaffold.  To this end, we performed IR
measurements with diffuse reflectance geometry using a liquid-N$_2$
cooled Indium Antimonide detector. Approximately 12 mg of sample was
mixed with 0.02~g of KBr and inserted into an atmospheric cell from
Thermo scientific. The sample was heated \emph{in situ} at 450~K in
vacuum for complete de-solvation. The pressure was kept constant and
changes were monitored as a function of time, while temperature was set
to 298~K. IR spectra were recorded every 10 seconds. 

By measuring the asymmetric stretch modes of both CO$_2$ and H$_2$O with
time, we find that typical MOF loading occurs in two consecutive steps.
At the beginning, the guest molecules condense in nano-pores on the
surface of the MOF, leading to very high-loading in those nano-pores.
From our experiments we conclude that this corresponds to approximately
9 molecules per unit cell, effectively clogging the pathways and
limiting diffusion of new approaching molecules~\cite{Nijem12}.  This
results in a red-shift (see supplementary materials), explained by
increasing lateral interactions between adsorbed
molecules~\cite{Nijem12}.  Our own calculations of the frequency shift
between low and high loading from above confirm that this shift
corresponds to high loading.  As time progresses, molecules diffuse from
highly-loaded pores to low-loading pores deeper inside the bulk of the
MOF. As a result, the asymmetric mode returns (with a blue-shift, see
supplementary material) to its original location. This peculiar
fluctuation in frequency-shifts is a clear signal of a two-states
molecular transport mechanism. In particular, the time it takes to
return to the original shift---i.e.\ to get from high to low
loading---is characteristic for the guest molecules and provides insight
into the diffusion process.

We have performed the above described diffusion experiment for CO$_2$
and H$_2$O, and find that it takes approximately 2 hour for water to
diffuse to the low-loading situation, while it takes only 22 minutes for
CO$_2$ (due to the difficulties of the experiments, we report these
times with a large error bar of approximately 25\%).  More importantly,
the ratio of those two times is
5.45. The described experimental situation clearly corresponds to
diffusion through mechanism c) and we can calculate the corresponding
ratio using the Arrhenius equation. To this end, we estimated the
pre-exponential diffusion factor using the harmonic transition-state
theory (see supplementary material)~\cite{Vineyard57}.  Using our
barrier heights from Table~\ref{table:bar} for CO$_2$ and H$_2$O for
mechanism c), we find for the same ratio a value of 5.43, in excellent
agreement with experiment.  This confirms that we have successfully
uncovered transport processes at the atomistic level, that govern the
transport processes at the experimentally measured macroscopic level.
Ideally, we would like to report other ratios, e.g.\ for H$_2$/CO$_2$,
but the corresponding IR experiments involving H$_2$ are very difficult
and results will be published in a longer, forthcoming article.

The fact that water diffuses so much slower than H$_2$ and CO$_2$ is to
be expected, but of paramount importance for practical applications of
MOFs. While the presence of water does not hinder the initial diffusion
of other gasses such as H$_2$ and CO$_2$ in the non-contaminated MOF
structure, high pressure or temperature can affect the delicate
thermodynamic equilibrium as incoming H$_2$O molecules displace
pre-adsorbed H$_2$ and CO$_2$ molecules.

In summary, we demonstrate that state of the art \emph{ab initio} NEB
simulations, coupled with time-resolved \emph{in situ} IR spectroscopy,
provide a complete atomistic understanding of the diffusion processes of
three important molecules in MOF-74-Mg for the purpose of hydrogen
storage and gas separation. Our theoretical atomistic model for the
molecular transport explains experimental IR macroscopic evidences.
Furthermore, our calculations clarify the two-state mechanism, observed
experimentally, which controls the macroscopic diffusion of these
molecules.  While the present study only focuses on MOF-74-Mg, it is our
belief that the same methodology can be successfully applied to unveil
the molecular transport in other MOFs and nano-porous materials,
providing further insight into the important question of diffusion as
well as a robust theoretical foundation to guide the interpretation of
challenging diffusion experiments.

This work was entirely supported by the Department of Energy Grant No.
DE-FG02-08ER46491.


\end{document}


\title{Diffusion of Small Molecules in\\
Metal Organic Framework Materials\\
--- Supplementary Materials ---}

\author{Pieremanuele Canepa}
\affiliation{Department of Physics, Wake Forest University,
Winston-Salem, NC 27109, USA.}

\author{Nour Nijem}
\affiliation{Department of Materials Science and Engineering, University
of Texas at Dallas, TX 75080, USA.}

\author{Yves J. Chabal}
\affiliation{Department of Materials Science and Engineering, University
of Texas at Dallas, TX 75080, USA.}

\author{T. Thonhauser}
\affiliation{Department of Physics, Wake Forest University,
Winston-Salem, NC 27109, USA.}

\date{\today}

\maketitle

\section{Details on the experimental measurements and observations}

\subsection{CO$_2$ Diffusion Experiment}

\begin{figure}
\includegraphics[width=0.9\columnwidth]{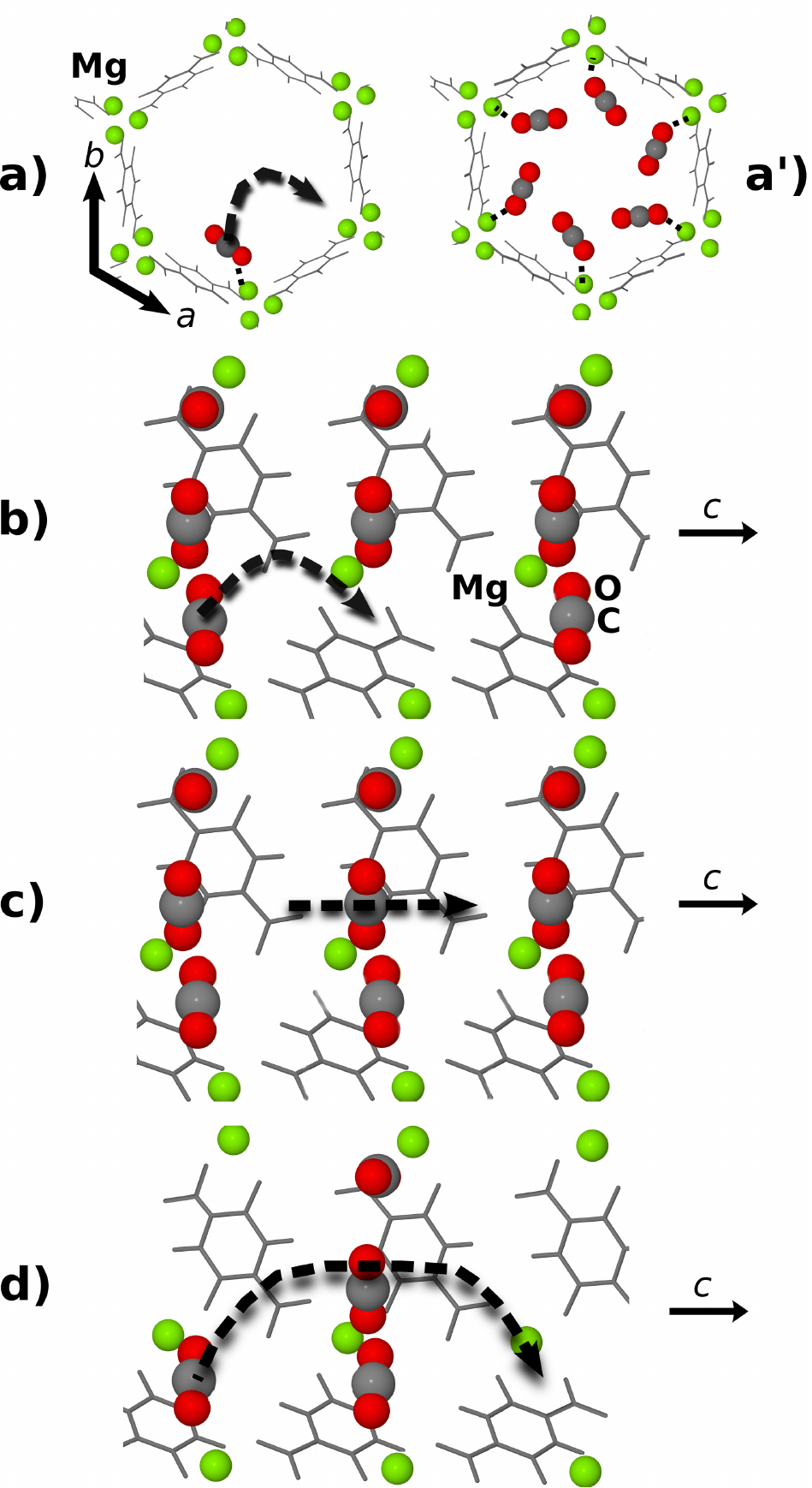}
\caption{\label{fig:exp1}Differential IR absorption spectra of the
asymmetric stretch mode of CO$_2$ adsorbed in MOF-74-Mg at 4.29 Torr.
Red (bottom) and blue (top) spectra show the initial (after 11 min.) and
final conditions (after 22 min.) after the addition of 0.29 Torr.}
\end{figure}

\noindent A specific amount of CO$_2$ gas was trapped and studied as a
function of time at a pressure of 4 Torr. After stabilization at that
pressure for one hour (no changes were observed), we perturbed this
state of equilibrium by adding a minute amount of pressure (0.29 Torr).
The \emph{differential} spectra in Fig.~\ref{fig:exp1} summarize the
dynamical changes of the CO$_2$ asymmetric stretching mode occurring at
4.29~Torr in a period of 22 minutes following the equilibration. We
report the differential spectra due to the otherwise overwhelming
absorption of the CO$_2$ gas present in the measurement chamber, so
that we can isolate the delicate dynamics involved in the diffusion of
this molecule in MOF-74-Mg. At the beginning (red spectrum) we observe
a red-shift in the asymmetric CO$_2$ stretch mode from its initial
value.  Although the red-shift occurs shortly after the increase of
the pressure, we show in Fig.~\ref{fig:exp1} the spectrum after 11
minutes because it takes some time to initiate the spectral
acquisition. Then, after 22 minutes this mode blue-shifts back to the
original value (blue spectrum). We report this time with a 25\% error
bar, as it is experimentally difficult to determine exactly when the
peak returns to its original position.  Figure~\ref{fig:exp1} is a
clear indication that diffusion occurs from the perturbed situation
with higher loading to a new equilibrium with lower loading after 22
minutes.  Note that in the differential spectra of
Fig.~\ref{fig:exp1}, the position of peaks in the derivative-looking
regions is affected by the absorption of the reference, leading to an
additional red-shift and blue-shift from the exact positions of the
peak. As such, from Fig.~\ref{fig:exp1} it is difficult to determine
absolute values for the true position and shifting of peaks.

However, from other recent IR absorption measurements we know that the
IR frequency of CO$_2$ adsorbed at the Mg metal center is at 2351
cm$^{-1}$ [\onlinecite{Nijem12}]. We further know that this frequency
red-shifts approximately $-$3~cm$^{-1}$ as the concentration of the
CO$_2$ increases, due to lateral CO$_2\cdots$CO$_2$ intermolecular
interactions [\onlinecite{Yao12,Nijem12}]. The observed red-shift (red
spectrum in Fig.~\ref{fig:exp1}) indicates a momentary increase in the
local CO$_2$ concentration. Then, as time progresses, the mode returns
(with a blue-shift) to it's original frequency, as the CO$_2$ molecules
diffuse from the higher loading situation to the lower loading situation
by occupying new available Mg sites deeper inside the MOF.  The
characteristic timescale for this diffusion is 22 minutes (with a 25\%
error bar).

\subsection{H$_2$O Diffusion Experiment}

\begin{figure}
\centering
\includegraphics[width=0.85\columnwidth]{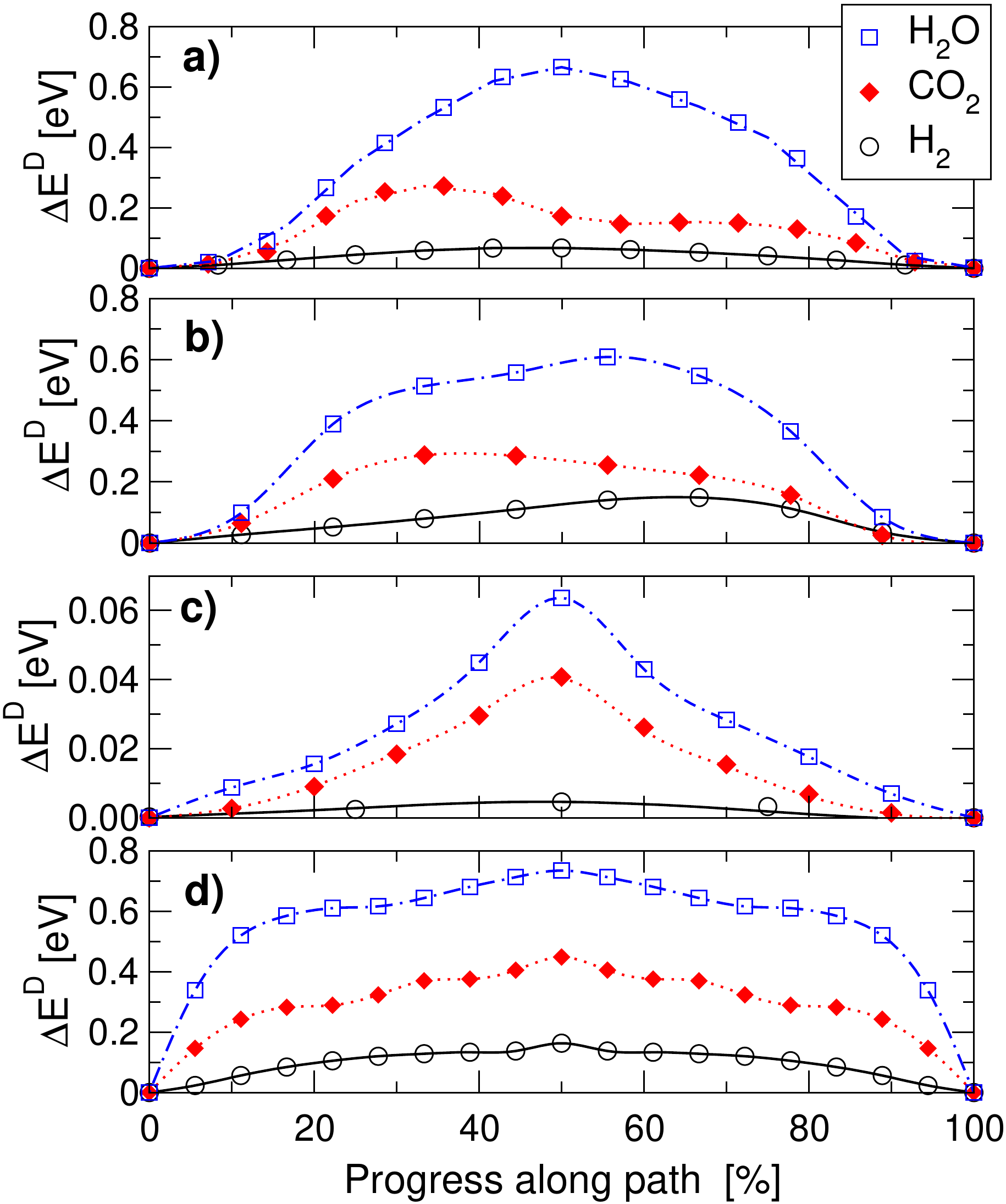}
\includegraphics[width=0.85\columnwidth]{Fig3}
\caption{\label{fig:exp2} IR absorption spectra of H$_2$O adsorbed in
MOF-74-Mg at a) 50 mTorr and b) 200 mTorr. The shaded area shows the
formation of hydrogen bonding. \emph{Sym.} and \emph{asym.} refer to
symmetric and asymmetric stretch modes of H$_2$O.}
\end{figure}

\noindent The diffusion of water was also studied at two pressures,
i.e.\ 50 mTorr and 200 mTorr. To this end, a known amount of water vapor
was introduced into an activated MOF-74-Mg and changes were monitored as
a function of time.  Figure~\ref{fig:exp2}  summarizes the IR absorption
spectra of the H$_2$O stretch modes after introducing 50~mTorr and
increasing the pressure to 200~mTorr. As the H$_2$O is first introduced
at 50~mTorr, two frequencies are observed at 3682 cm$^{-1}$ and 3586
cm$^{-1}$ and are assigned to the asymmetric and symmetric stretching
O--H modes of water molecules adsorbed at the Mg metal sites.  When
increasing the pressure to 200~mTorr, we observe that the asymmetric
stretch red-shifts down to 3667 cm$^{-1}$ (a red-shift of --15
cm$^{-1}$). However, more importantly, under this higher pressure we
find a broad absorption in the region between 3000 cm$^{-1}$ and 3500
cm$^{-1}$, which shows the presence of hydrogen-bonded water,
characteristic of a high-loading situation. We find that the asymmetric
stretch mode returns to its original value after a characteristic time
of 2 hours. Again, we report this number with a 25\% error bar.

The overall situation is very similar to the case of CO$_2$ from above:
Initially, the increase of pressure causes temporarily an increase in
loading, which---after a characteristic timescale---returns to a
low-loading situation by means of diffusion.  During this diffusion,
CO$_2$ has to overcome a smaller barrier than H$_2$O in order to
penetrate deep into the MOF structure, clearly evident from the
different timescales required to reach the new equilibrium low-loading
situation. In particular, we find for the ratio of their characteristic
timescales.
\begin{equation}\label{eq:ratio}
\tau = \frac{t_{\rm H_2O}}{t_{\rm CO_2}}  = \frac{120 \;\rm min}{22\;\rm
min} = 5.45\;.
\end{equation}

Although the IR experiment shows the occurrence of barrier-like
diffusion phenomena for both CO$_2$ and H$_2$O, the estimation of the
exact time resolution of these processes becomes extremely delicate due
to the complexity of the experimental measurements. The large difference
in the timescale, minutes for CO$_2$ and hours for H$_2$O, clearly
illustrates the presence of ongoing changes to be attributed to a change
in the local environment experienced by the guest molecules.

\section{Connection between measured diffusion times and calculated
barrier heights}

\noindent The experimental time ratio $\tau$ from Eq.~(\ref{eq:ratio})
can be linked to the ratio of our calculated NEB diffusion barriers for
CO$_2$ and H$_2$O. In particular, we can write
\begin{equation} \label{eq:ratio2} \tau = \frac{t_{\rm H_2O}}{t_{\rm CO_2}}
\simeq \frac{D_{\rm CO_2}}{D_{\rm H_2O}} = \frac{D^0_{\rm CO_2}}{D^0_{\rm
H_2O}}\: \exp\Big[-\frac{\Delta E^D_{\rm CO_2}
- \Delta E^D_{\rm H_2O} }{k_b T} \Big]\;, \end{equation}
where $D_x$ is the diffusion rate, $D^0_x$ is the pre-exponential factor
that enters the empirical Arrhenius equation, $\Delta E^D_x$ is the NEB
diffusion barrier. The pre-exponential factors were approximated using
the harmonic transition-state theory [\onlinecite{Vineyard57}], starting
from the harmonic frequencies computed for the diffusing molecules. Each
pre-exponential factor was calculated via $\prod _i^{N}\nu^{(i)}_{\rm
gas}/\prod _i^{N-1}\nu^{(i)}_{\rm ads}$, where $\nu_{\rm gas}$ and
$\nu_{\rm ads}$ are the harmonic frequencies of the molecules when in
gas-phase and adsorbed, respectively. For the ratio of the
pre-exponential of Eq.~(\ref{eq:ratio2}) we thus get $D^0_{\rm
CO_2}/D^0_{\rm H_2O}=2.49$.  Using a temperature of 298 K (experimental
condition), for the exponential piece we get $\exp[-(\Delta E^D_{\rm
CO_2} - \Delta E^D_{\rm H_2O})/k_b T]=2.18$, such that overall we find
$\tau = 2.49\times 2.18 = 5.43$, in excellent agreement with experiment.
Note that this high level of agreement is probably to some extent
accidental, since the calculation of the pre-exponential factor involves
several substantial approximations. In addition, the experimental
timings are subjected to fluctuations imposed by the complexity of the
experimental measurements.

